\documentclass{article}
\usepackage{spconf,amsmath,graphicx}
\usepackage{xcolor}
\usepackage{hyperref}

\usepackage{multirow}

\title{Learning to Restore Images Degraded by Atmospheric\\ Turbulence using Uncertainty}
\name{Rajeev Yasarla and Vishal M. Patel\thanks{This research was supported by NSF CAREER award 2045489.}}
\address{Department of Electrical \& Computer Engineering\\ Johns Hopkins University\\
\{ryasarl1, vpatel36\}@jhu.edu}
\graphicspath{{Figs/}}
\begin{document}
%
\maketitle
\begin{abstract}
	Atmospheric turbulence can significantly degrade the quality of images acquired by long-range imaging systems by causing spatially and temporally random fluctuations in the index of refraction of the atmosphere. Variations in the refractive index causes the captured images to be geometrically distorted and blurry. Hence, it is important to compensate for the visual degradation in images caused by atmospheric turbulence.  In this paper, we propose a deep learning-based approach for restring a single image degraded by atmospheric turbulence.  	We make use of the epistemic uncertainty based on Monte Carlo dropouts to capture regions in the image where the network is having hard time restoring.  The estimated uncertainty maps are then used to guide the network to obtain the restored image.   Extensive experiments are conducted on synthetic and real images to show the significance of the proposed work.  Code is available at : \href{https://github.com/rajeevyasarla/AT-Net}{https://github.com/rajeevyasarla/AT-Net}

\end{abstract}
\section{Introduction}
Atmospheric turbulence can significantly degrade the quality of images acquired by long-range visible and thermal imaging systems by causing spatially and temporally random fluctuations in the index of refraction of the atmosphere. Variations in the refractive index causes the captured images to be geometrically distorted and blurry \cite{Turbulence_Book,tatarski2016wave}. Hence, it is very important to compensate for the visual degradation in images caused by atmospheric turbulence. Adaptive optics-based techniques can be used to compensate for turbulence effects in images. However, they require very large, complex, fragile, and expensive hardware. On the other hand, image processing-based approaches are cheap and effective.  In this paper, we pose the turbulence degraded image restoration problem as a nonlinear regression problem, where the optimal parameters are learned from synthetically generated data. As a function approximator, we propose to use Deep Convolutional Neural Networks (DCNNs).

Under the assumption that the scene and the imaging sensor are both static and that observed motions are due to the air turbulence alone, the image degradation due to atmospheric turbulence can be mathematically formulated as follows \cite{TurbulenceModelFrakes,furhad2016restoring, zhu2012removing, hirsch2010efficient,Variational_turbulencee}
\begin{equation}	\label{eq:model}
\mathbf{y}_{k} = \mathbf{G}_{k}(\mathbf{H}_{k}(\mathbf{x})) + \boldsymbol{\epsilon}_{k}, \;\;\; k=1, \cdots, M,
\end{equation}		
where $\mathbf{x}$ denotes the ideal image, $\mathbf{y}_{k}$ is the $k$-th observed frame, $\mathbf{G}_{k}$ and $\mathbf{H}_{k}$ represent the deformation operator and air turbulence-caused blurring operator, respectively and $\boldsymbol{\epsilon}_{k}$ denotes additive noise.  The deformation operator is assumed to deform randomly and $\mathbf{H}_{k}$ correspond to a space-invariant diffraction-limited point spread function (PSF).  As can be seen from \eqref{eq:model}, atmospheric turbulence has two main degradations on the observed images:  geometric distortion and blur.  Various optics-based \cite{pearson1976atmospheric,tyson2015principles,Turbulence_Book} and image processing-based \cite{metari2007new,shimizu2008super,furhad2016restoring,meinhardt2014implementation,micheli2014linear,zhu2012removing,lau2019restoration,chak2018subsampled,Variational_turbulencee, PCA_Turbulence_restoration,lau2020atfacegan} turbulence removal algorithms have been proposed in the literature.  In general, most image processing methods follow a similar pipeline: lucky region fusion or non-rigid image registration, and blind deconvolution.

\begin{figure}[t!]
	\begin{center}
		\centering
		\includegraphics[width=0.1\textwidth]{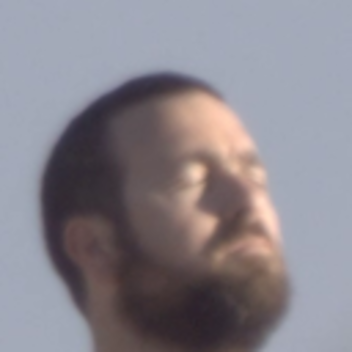}
		\includegraphics[width=0.1\textwidth]{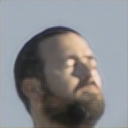}
		\includegraphics[width=0.1\textwidth]{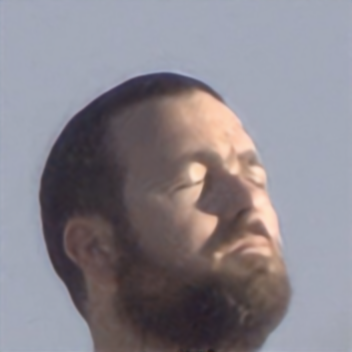}\\
		Input\hskip25ptTDRN \cite{yasarla2020learning}\hskip25pt AT-Net
	\vskip-10pt	\caption{Sample image restoration results comparing TDRN with the proposed metod, AT-Net.}
		\label{Fig:face}
	\end{center}
\end{figure}

\begin{figure*}[t!]
	\begin{center}
		\centering
		\includegraphics[width=.85\textwidth]{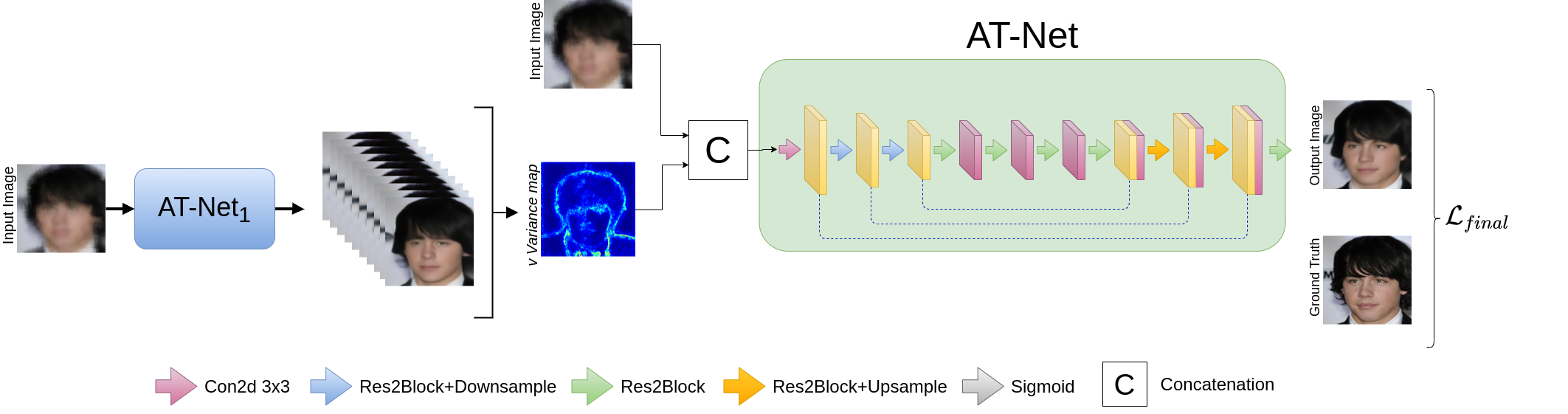}
		\vskip-10pt\caption{An overview of the proposed AT-Net atmospheric turbulence degraded image restoration method.}
		\label{Fig:TDRN}
	\end{center}
\end{figure*}
 In many applications of long-range imaging, such as surveillance, we are faced with a scenario where we have to restore a single image degraded by atmospheric turbulence.  In this case, we have the following observation model, 
\begin{equation}	\label{eq:general_model}
\mathbf{y} = \mathbf{G}(\mathbf{H}(\mathbf{x})) + \boldsymbol{\epsilon},
\end{equation}	
where the subscript $k$ has been removed from \eqref{eq:model}.  This is an extremely ill-posed problem as we have to overcome the effects of both blur and geometric distortion from a single image.  Our prior work \cite{yasarla2020learning} was one of the first deep learning-based attempts in the literature to recover a single image degraded by atmospheric turbulence.   In \cite{yasarla2020learning} we first estimate the prior information  regarding  the  amount  of  geometric  distortion  and blur at each pixel using two separate networks based on Monte Carlo dropouts \cite{kendall2017uncertainties}. The estimated priors are then used by  a  restoration network  called,  Turbulence  Distortion  Removal  Network (TDRN),  to  restore an image.  Although   this  framework  is  capable  of  alleviating blur and geometric distortion caused by atmospheric turbulence, it requires the design and optimization of three separate networks --  image   deblurring   network to estimate the blur prior,  geometric   distortion   removal   network to estimate the distortion prior and TDRN to restore an image.   

Rather than dealing with geometric distortion and blurry degradation separately, in this paper we improve our work in \cite{yasarla2020learning} by developing an approach that estimates only a single prior (in terms of uncertainty) corresponding to a combination of both blur and geometric distortions in turbulence degraded images.   The estimated uncertainty maps are then used to guide the network to obtain the restored image.   Extensive experiments are conducted on synthetic and real images to show the significance of the proposed work.  Figure~\ref{Fig:face} presents a comparison between TDRN and the improved version of TDRN proposed in this paper, which we call AT-Net (Atmospheric Turbulence distortion removal Network).  As can be seen from this figure, the new approach is able to recover details in the image better than \cite{yasarla2020learning}.

\section{Proposed Method}
\begin{figure}[t!]
	\begin{center}
		\centering
		\includegraphics[width=0.11\textwidth]{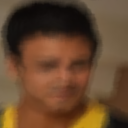}
		\includegraphics[width=0.11\textwidth]{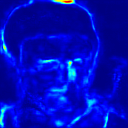} 
		\includegraphics[width=0.11\textwidth]{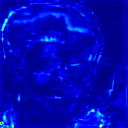}
		\includegraphics[width=0.11\textwidth]{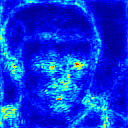}\\ \vskip3pt
		\includegraphics[width=0.11\textwidth]{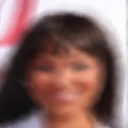}
		\includegraphics[width=0.11\textwidth]{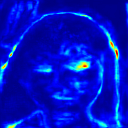} 
		\includegraphics[width=0.11\textwidth]{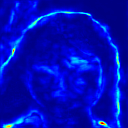}
		\includegraphics[width=0.11\textwidth]{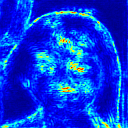}\\ 
		(a) \hskip45pt(b)\hskip45pt(c)\hskip45pt(d)
	\end{center}
	\vskip-20pt\caption{Image priors in terms of uncertainty maps on sample turbulence distorted images. (a) Turbulence degraded images.  (b) Blur prior \cite{yasarla2020learning}. (c) Geometric distortion prior \cite{yasarla2020learning}. (d) Distortion map estimated using the proposed method.}
	\label{Fig:var_maps}
	\vskip-15pt
\end{figure}

Figure~\ref{Fig:TDRN} gives an overview of the proposed restoration network.  Given pairs of clean and atmospheric turbulence distorted images, $\{\mathbf{y}_{k},\mathbf{x}_{k}\}_{k=1}^{N}$, we  train two networks similar to AT-Net for restoration.  We refer to them as $\text{AT-Net}_{1}$ and AT-Net.  By applying Monte Carlo dropout in every layer of $\text{AT-Net}_{1}$, we can formulate the epistemic uncertainty \cite{kendall2017uncertainties} and use the corresponding variance as a prior information.  For example, given a turbulence distorted image $\mathbf{y}$, we pass it as an input to $\text{AT-Net}_{1}$ $S$ times and obtain a set of outputs $\{\mathbf{p}_i\}_{i=1}^S$, where $i$ corresponds to the $i$th instance of dropout with corresponding parameters $\boldsymbol{\theta}^{i}$ and $\mathbf{p}_i = \text{AT-Net}_{1}(\mathbf{y};\boldsymbol{\theta}^{i})$. We define the distortion prior $\mathbf{d}$ as the variance of the outputs $\{\mathbf{p}_i\}_{i=1}^S$, i.e. $\mathbf{d} = \text{variance}(\{\mathbf{p}_i\}_{i=1}^S)$. As explained in \cite{kendall2017uncertainties} this variance is defined as the model uncertainty. However, in our case we use it as a measure of the ability or competence of  the network in addressing image restoration. Hence a high variance value at a pixel location in $\mathbf{d}$ means that the restoration network is not able to reconstruct the underlying clean image properly at that pixel  in the output image.  Figure~\ref{Fig:var_maps} compares the variance maps estimated by the proposed method with blur and geometric distortion variance maps estimated by our previous work \cite{yasarla2020learning}.  We can clearly see that the combined prior closely resembles the degradations that are present in the input image compared to the blur and geometric deformation priors.\\

\noindent {\bf{AT-Net$_1$. }}  The goal of the AT-Net$_1$ network is to estimate the uncertainty maps corresponding to the input degraded images.  To this end, AT-Net$_1$ is trained to restore degraded images.  In particular,  AT-Net$_1$ is constructed using the  UNet~\cite{ronneberger2015u}  architecture with Res2Block as the basic building block \cite{gao2019res2net}. AT-Net$_1$ consists of the following layers,\\
Res2Block(3,64)-Downsample-Res2Block(64,64)-Downsample-Res2Block(64,64)\\
-Res2Block(64,64)-Res2Block(64,64)-Res2Block(64,64)-Res2Block(64,64)\\
-Upsample-Res2Block(64,64)-Upsample-Res2Block(64,16)-Res2Block(16,3),\\
where Downsample means average pooling layer and  Res2Block $(m,n)$ denotes Res2Block with $m$ input channels and $n$ output channels.\\ 

\noindent {\bf{AT-Net. }}  Given turbulence degraded image $\mathbf{y}$,  turbulence variance map $\mathbf{d}$ is first computed using AT-Net$_1$.  Then, $\mathbf{y}$ and $\mathbf{d}$ are used as inputs to AT-Net to obtain the restored image $\hat{\mathbf{x}}$, \textit{i.e} $\hat{\mathbf{x}}=$ AT-Net$(\mathbf{x},\mathbf{d})$.  The AT-Net network architecture consists of the following layers,
Conv2d $3\times 3$(5,16)- Res2Block(16,64)- Downsample\\Res2Block(64,64)-Downsample-Res2Block(64,64)\\-Res2Block(64,64)-Res2Block(64,64)
-Res2Block(64,64)\\-Res2Block(64,64)-Upsample-Res2Block(64,64)\\-Upsample-Res2Block(64,3),\\
where Conv2d $3\times 3(m,n)$ denotes a $3\times 3$ convolutional layer with $m$ input channels and $n$ output channels.

 The loss function we use to train both AT-Net$_1$ and AT-Net networks is a combination of the perceptual loss and the $\mathcal{L}_1$ loss
\begin{equation}\mathcal{L} = \mathcal{L}_1 + \lambda_p \mathcal{L}_p,
\end{equation}
 where  $\mathcal{L}_1 = \|\hat{\mathbf{x}}-\mathbf{x}\|_1$ and $\mathcal{L}_p = \| F(\hat{\mathbf{x}})-F(\mathbf{x})\|_2^{2}.$  Here, $F(.)$ denotes the features from layer $pool3$ of a pretrained VGG-Face network \cite{parkhi2015deep}.  In our experiments, we set $\lambda_p$ equal to 0.002.

\section{Experimental Results}
In this section, we evaluate the proposed AT-Net method on both synthetic and real-world datasets \cite{yasarla2020learning}.  The performance of different methods on the synthetic data is evaluated in terms of Peak Signal-to-Noise Ratio (PSNR), Structural Similarity Index (SSIM) and $d_{VGG}$.  Here, $d_{VGG}$ is defined as feature distance between the restored image $\hat{\mathbf{x}}$ and the ground truth clean image $\mathbf{x}$.  We use the outputs of $pool5$ layer from the VGG-Face~\cite{parkhi2015deep} to compute $d_{VGG}$. 
Furthermore, in order to show the significance of different face restoration methods, we perform face recognition on the restored images using ArcFace \cite{deng2019arcface}.    The performance of the proposed AT-Net method is compared against the following recent state-of-the-art face image restoration methods \cite{pan2014deblurring,shen2018deep,yasarla2019deblurring} and generic single image deblurring methods \cite{kupyn2018deblurgan,zhang2019deep}.  Note that we re-train these methods using the same turbulence degraded images that are used to train our network.  While re-training the networks, we followed the training procedure including the parameter selection mentioned in those respective papers.\\
\noindent {\bf{Training Details. }}  We collected 2,000 clean Face images for training AT-Net  from the Helen dataset~\cite{le2012interactive}.  In addition, 25,000 images were randomly selected from the CelebA dataset~\cite{liu2015deep}.  Atmospheric turbulence degraded images are generated using the turbulence synthesis approach presented in ~\cite{chak2018subsampled,yasarla2020learning}. As a result, we generated 1.5 million pairs (i.e. $\{\mathbf{y}_{i}, \mathbf{x}_{i}\}_{i=1}^{1.5\times 10^{6}}$) of turbulence distorted images and corresponding clean images for training the AT-Net$_1$ and AT-Net networks.  Given $\mathbf{y}$, we estimate $\mathbf{d}$ (turbulence variance-map) using AT-Net$_{1}$ where we set $S=10$.  Adam optimizer with learning rate of $2\times10^{-4}$ and a batch-size of 10 is used for training both networks. We train AT-Net$_{1}$ and AT-Net for $2 \times 10^5$ and $1.5\times 10^{6}$ iterations, respectively.\\
\noindent {\bf{Testing. }} The synthetic test set contains 24,000 images from Helen and 24,000 images from CelebA  as test images (i.e. in total 48,000 test images denoted by $\mathcal{D}_{test}$).  In addition, we use 600 real-world turbulence distorted images collected by the US Army in a variety of different atmospheric conditions as another test set.

\begin{table}[h!]
	\caption{Quantitative results in terms of PSNR, SSIM, and $d_{VGG}$ on the synthetic datasets. PSNR/SSIM higher the better, and $d_{VGG}$ lower the better} \label{Table:CompSt}
	\centering 
	\resizebox{0.48\textwidth}{!}{
		\begin{tabular}{|l|c|c|c|c|c|c|}
			\hline
			\multirow{2}{*}{\begin{tabular}[c]{@{}c@{}}Deturbulence\\ Method\end{tabular}} & \multicolumn{3}{c|}{CelebA}   & \multicolumn{3}{c|}{Helen}   \\ \cline{2-7} 
			& PSNR   & SSIM    & $d_{VGG}$   & PSNR   & SSIM    & $d_{VGG}$   \\ \hline
			Turbulence-distorted   &  22.43    &    0.731   &   5.13      &   22.35     &   0.667     &  6.11       \\ \hline
			\multicolumn{1}{|l|}{Pix2Pix~\cite{isola2017image}(CVPR 2017)}   &  22.51 &  0.738  & 5.28 & 22.62    &  0.671    & 5.83 \\ \hline
			Pan et al.~\cite{pan2014deblurring}(ECCV 2014)    &   20.73    &   0.679  &  6.04    &    20.01    &   0.627    &  7.28  \\ \hline
			Shen et al.~\cite{shen2018deep}(CVPR 2018)   &   23.08      &    0.745 &  4.72   &  23.01  &   0.681 &  5.14 \\ \hline
			Yasarla et al.~\cite{yasarla2019deblurring}(TIP 2020)    & 24.06 &   0.768   &    4.11    &    23.81      &  0.702       &   4.49       \\ \hline
			Kupyn et al.~\cite{kupyn2018deblurgan}(CVPR 2018)      &   23.54     &   0.748    &   4.51   &   23.28   &   0.693     &  4.98   \\ \hline
			Zhang et al.~\cite{zhang2019deep}(CVPR 2019)     &  24.16    &   0.770 &  3.94     &   23.95    &    0.709    &   4.36     \\ \hline
			\multicolumn{1}{|l|}{TDRN\cite{yasarla2020learning} (trained with $\mathcal{L}_1, \mathcal{L}_p$)}  & 25.04   &    0.802      &   3.57     &    24.47      &   0.725     &    4.17 \\ \hline
			AT-Net (ours) &      \textbf{25.31} &    \textbf{0.810}  &  \textbf{3.09}    &  \textbf{24.95}   &   \textbf{0.750}  & \textbf{3.80} \\ \hline
	\end{tabular}}
\end{table}

\begin{table}[hp!]
	\caption{Top-1, Top-3 and Top-5 face recognition accuracies on a real-world dataset.} \label{Table:Real_comp}
	\centering
	\resizebox{0.4\textwidth}{!}{
		\begin{tabular}{|l|c|c|c|}
			\hline
			\begin{tabular}[c]{@{}c@{}} Method\end{tabular} & Top-1                 & Top-3                 & Top-5                 \\ \hline
			Turbulence-distorted         &  38.10  &   49.32   &  56.13   \\ \hline
			\multicolumn{1}{|l|}{Pix2Pix~\cite{isola2017image}(CVPR 2017)}       &   37.82     &   50.76    &  57.41    \\ \hline
			Pan et al.~\cite{pan2014deblurring} (ECCV 2014)      & 35.67    & 45.18         & 50.79        \\ \hline
			Shen et al.~\cite{shen2018deep}(CVPR 2018)       & 39.91   &    52.21      &   58.17  \\ \hline
			Yasarla et al.~\cite{yasarla2019deblurring}(TIP 2020)       &  42.31      &   57.72    &   64.40   \\ \hline
			Kupyn et al.~\cite{kupyn2018deblurgan}(CVPR 2018)    & 40.72   &  54.86   &   62.28   \\ \hline
			Zhang et al.~\cite{zhang2019deep}(CVPR 2019)       &  44.76     &   60.96    &    69.75    \\ \hline
			\multicolumn{1}{|l|}{TDRN\cite{yasarla2020learning} (trained with $\mathcal{L}_1, \mathcal{L}_p$)}  & 47.16 &  62.91 & 72.87\\ \hline
			AT-Net (ours)  &   \textbf{48.38}     &   \textbf{63.96}    &  \textbf{73.83}    \\ \hline
		\end{tabular}}
		\vskip-10pt
\end{table}

\begin{figure*}[htp!]
	\begin{center}
		\centering
		\includegraphics[width=0.115\textwidth]{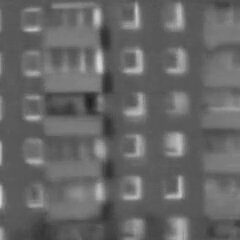}
		\includegraphics[width=0.115\textwidth]{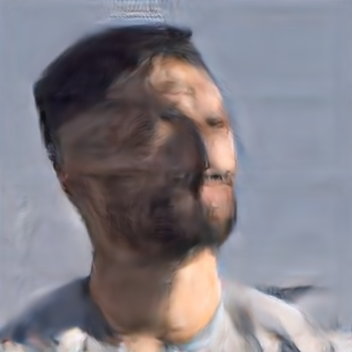} 
		\includegraphics[width=0.115\textwidth]{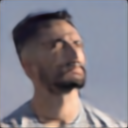}
		\includegraphics[width=0.115\textwidth]{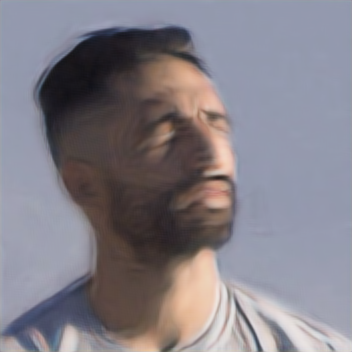}
		\includegraphics[width=0.115\textwidth]{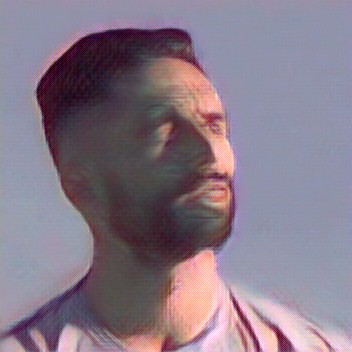}
		\includegraphics[width=0.115\textwidth]{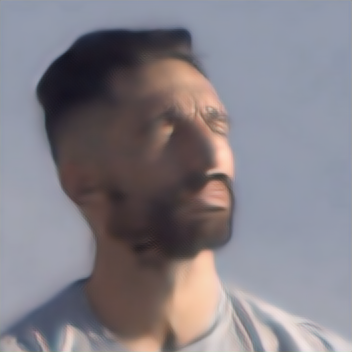}
		\includegraphics[width=0.115\textwidth]{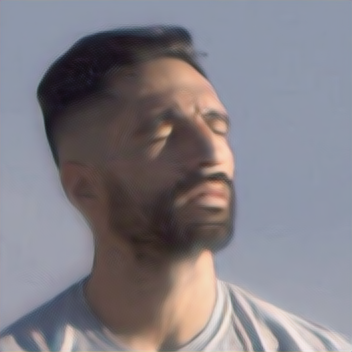}\
		\includegraphics[width=0.115\textwidth]{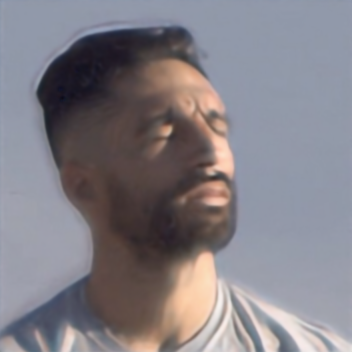}\ \vskip3pt
		\includegraphics[width=0.115\textwidth]{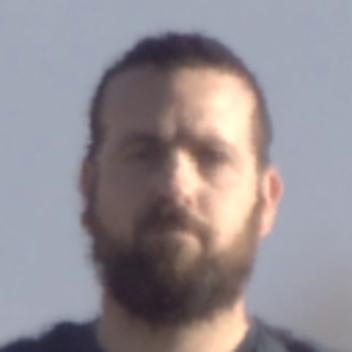}
		\includegraphics[width=0.115\textwidth]{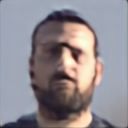} 
		\includegraphics[width=0.115\textwidth]{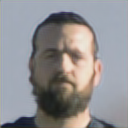}
		\includegraphics[width=0.115\textwidth]{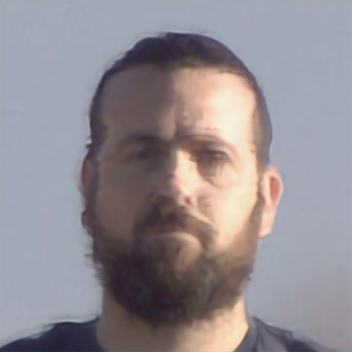}
		\includegraphics[width=0.115\textwidth]{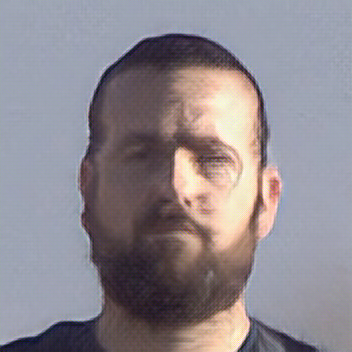}
		\includegraphics[width=0.115\textwidth]{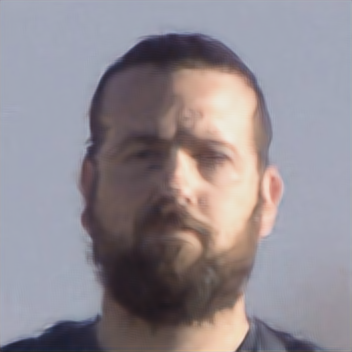}
		\includegraphics[width=0.115\textwidth]{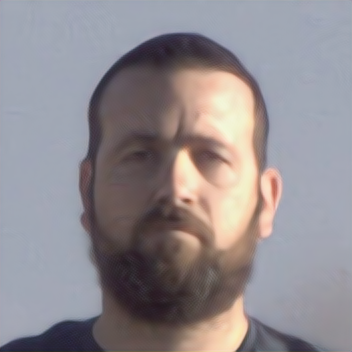}
		\includegraphics[width=0.115\textwidth]{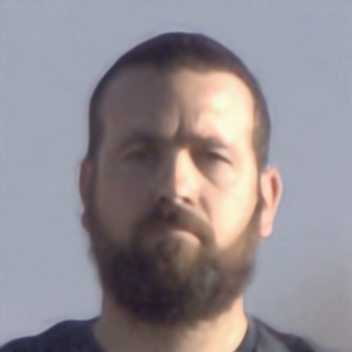}\\ 
		\vskip3pt
		\includegraphics[width=0.115\textwidth]{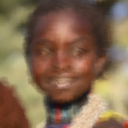}
		\includegraphics[width=0.115\textwidth]{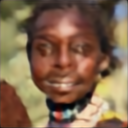} 
		\includegraphics[width=0.115\textwidth]{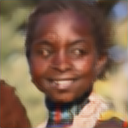}
		\includegraphics[width=0.115\textwidth]{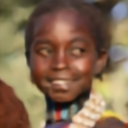}
		\includegraphics[width=0.115\textwidth]{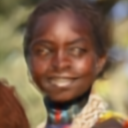}
		\includegraphics[width=0.115\textwidth]{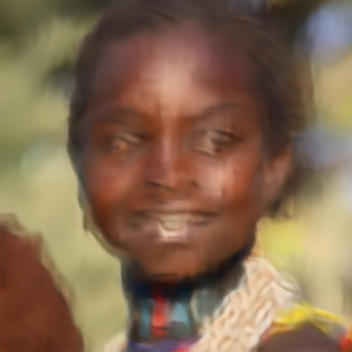}
		\includegraphics[width=0.115\textwidth]{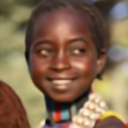}
		\includegraphics[width=0.115\textwidth]{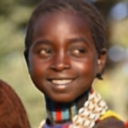}\\ 
		\vskip3pt
		\includegraphics[width=0.115\textwidth]{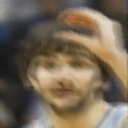}
		\includegraphics[width=0.115\textwidth]{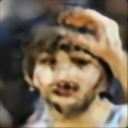} 
		\includegraphics[width=0.115\textwidth]{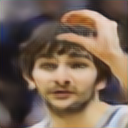}
		\includegraphics[width=0.115\textwidth]{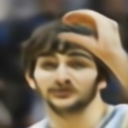}
		\includegraphics[width=0.115\textwidth]{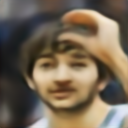}
		\includegraphics[width=0.115\textwidth]{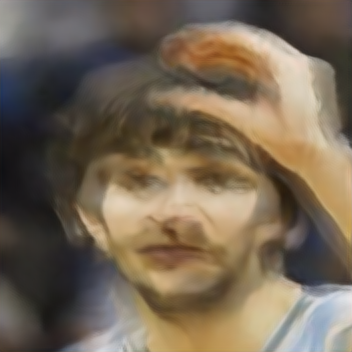}
		\includegraphics[width=0.115\textwidth]{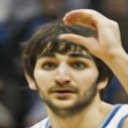}
		\includegraphics[width=0.115\textwidth]{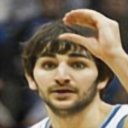}\\ 
		Input\hskip50pt\cite{pan2014deblurring}\hskip45pt\cite{shen2018deep}\hskip40pt\cite{yasarla2019deblurring}\hskip40pt\cite{kupyn2018deblurgan}\hskip45pt\cite{zhang2019deep}\hskip40pt\cite{yasarla2020learning}\hskip40pt AT-Net\\
		\vskip-12pt 
		\caption{Image restoration results on sample real-world (rows 1-2) and synthetic (rows 3-4) turbulence distorted images. Compared to the other methods, the proposed AT-Net method produces sharp and clean face images.}
		\label{Fig:quali_cmp}
	\end{center}
	\vskip-25pt
\end{figure*}

\noindent {\bf{Results on Synthetic Data. }}  Results corresponding to different methods on synthetic data are shown in  Table~\ref{Table:CompSt} and Fig~\ref{Fig:quali_cmp} (Rows 3 \& 4). Higher PSNR/SSIM  and lower $d_{VGG}$ correspond to better quality of the  reconstructed images.  As can be seen from Fig~\ref{Fig:quali_cmp} (Rows 3 \& 4) and  Table~\ref{Table:CompSt}, AT-Net outperforms the state-of-the-art face image restoration methods.  In particular, generic  deblurring methods~\cite{kupyn2018deblurgan,zhang2019deep} are not able to perform well due to the lack of prior information.  On the other hand,  methods that make use of some prior information about the face \cite{pan2014deblurring,shen2018deep,yasarla2019deblurring} are unable perform better because of improper prior estimation from the input images. AT-Net outperforms the state-of-the-art methods including our prior method \cite{yasarla2020learning}, which clearly demonstrates the effectiveness of the proposed method.\\
\noindent {\bf{Results on Real-world Data. }}  We also evaluate the performance of different methods on  several real-world turbulence distorted images collected by the US Army, published in our previous work~\cite{yasarla2020learning}.  Fig~\ref{Fig:quali_cmp} (Rows 1 \& 2) illustrate the qualitative performance of different methods on two sample real-world turbulence distorted face images from this dataset.  As can  be  seen  from  this  figure,  state-of-the-art restoration methods produce artifacts and blurry outputs especially around the mouth, eyes, and nose regions of the face. On the other hand, AT-Net is able to recover details of the face better and significantly improves the visual quality. 
\begin{figure}[h!]
	\begin{center}
		\centering
		\includegraphics[width=0.155\textwidth]{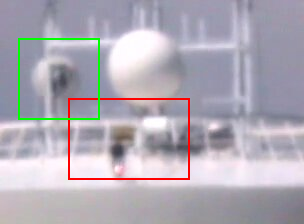}
		\includegraphics[width=0.155\textwidth]{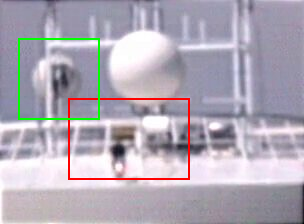}
		\includegraphics[width=0.155\textwidth]{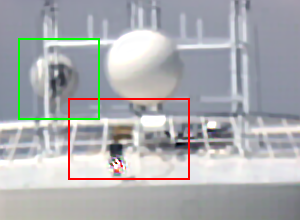}\\
		\includegraphics[height=1.07cm]{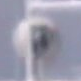}
		\includegraphics[height=1.07cm]{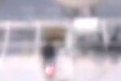}
		\includegraphics[height=1.07cm]{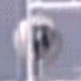}
		\includegraphics[height=1.07cm]{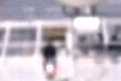}
		\includegraphics[height=1.07cm]{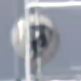}
		\includegraphics[height=1.07cm]{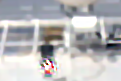}\\
		Input\hskip60pt\cite{yasarla2020learning}\hskip60pt AT-Net
		\vskip-10pt	\caption{Image restoration results on sample real-world turbulence distorted non-face images}
		\label{Fig:non_face}
	\end{center}
	\vskip-25pt
\end{figure}

Additionally, we perform face recognition task on real-world turbulence distorted images using ArcFace~\cite{deng2019arcface}.  The most similar faces (i.e. Top-K nearest matches) from the restored image are selected from the gallery set (set of different clean images corresponding to 100 identities) to check whether they belong to the same identity or not. Results corresponding to this experiment are shown in Table~\ref{Table:Real_comp}.  As can be seen from this table, AT-Net is able to restore real-world images better and preserves identity in  the restored images better than the other methods.  In particular, AT-Net gives more than 1\% improvement in Rank 1, 3 and 5 accuracy compared to our previous method \cite{yasarla2020learning}.\\
\noindent {\bf{Results on Real-world Non-face Images. }} Finally, we conducted experiments on non-face real-world atmospheric-turbulence degraded images to show that our  method can be used to restore non-face images as well.  As can been seen from  Fig~\ref{Fig:non_face},  AT-Net is able to restore better quality images as compared to our previous work ~\cite{yasarla2020learning}.

\section{Conclusion}
We proposed AT-Net to address the atmospheric turbulence distortion removal problem from a single image. This work can be viewed as an extension of our previous approach \cite{yasarla2020learning}.  Rather than having two separate networks for estimating blur and geometric deformation uncertainties, we only use a single network to estimate the overall distortion introduced by turbulence.  As a result, the proposed method reduces the complexity and at the same time is able to provide on par or better results compared to previous state-of-the-art methods. We conducted extensive experiments using synthetic and real data to show that  AT-Net network  mitigates the geometric deformation and blur introduced by turbulence, and restores better quality  images.

\bibliographystyle{IEEEbib}
{\small{
\bibliography{refs}}}

\end{document}